\documentclass[aps,prl,twocolumn,groupedaddress]{revtex4}
\usepackage{epsfig}
\usepackage{amsmath}
\usepackage{amsfonts}

\begin{document}


\title{Negative refraction by a virtual photonic lattice}
\author{Neil V. Budko and Shreyas B. Raghunathan}
\affiliation{Laboratory of Electromagnetic Research, Faculty of Electrical Engineering, Mathematics and Computer Science,
Delft University of Technology,
Mekelweg 4, 2628 CD Delft, The Netherlands}
\email{n.v.budko@tudelft.nl}



\date{\today}

\begin{abstract}
Research on photonics and metamaterials constantly challenges our intuitive understanding of the behaviour of light. In recent years we have seen negative refraction, focusing of light by a flat slab, a ``perfect'' prism, and an ``invisibility cloak'' \cite{1,2,3,4,5,6}. It is generally understood that the cause of this unusual behaviour is the strong (anomalous) dispersion, i.e., dependence of the material properties on the frequency of light. Dispersion can be either due to a natural microscopic resonance of the material as with surface plasmons-polaritons, or due to an effective resonance (band-gap) of the periodic lattice as in photonics \cite{7,8,9}. Metamaterials take the better of the two approaches representing a periodic array of designer subwavelength particles tuned to resonate at a specific frequency-band. At present, however, we have only a very basic understanding of the effect which a finite size of a sample of a periodic photonic crystal or metamaterial has on the macroscopic properties such as refraction. Yet every finite dielectric object is a moderate-quality resonator whose eigenmodes form a virtual photonic lattice with its own angular band-gaps and preferred directions of propagation. Here we show that this virtual lattice produces nontrivial real effects and that even a homogeneous dielectric resonator may refract negatively without either negative or periodically modulated permittivity/permeability. We also propose a simple way to control the period of this virtual photonic lattice by varying the transverse dimension of the resonator. Our research shows the importance of three-dimensional resonant phenomena in optics and may result in new optical devices with unusual properties. \end{abstract}


\maketitle
Unlike cavity resonators with perfectly conducting walls, three-dimensional rectangular dielectric resonators cannot be studied analytically, and we do not know the precise from of their eigenmodes or the values of the corresponding complex eigenfrequencies. All we have is a few mathematical estimates and a basic phenomenological understanding gained from numerical simulations \cite{10,11,12,13,14}. It is clear though that the quality of such resonators is never too high, as there are no eigenvalues equal to zero \cite{12,13}. Some of the eigenvalues may be close to zero, though, meaning that the associated eigenmode is close to resonance and will be relatively amplified. Inside the object a spatial eigenmode has the form of a ``frozen'' standing wave pattern being the result of interference due to reflections from the boundaries of the object. If several eigenmodes are close to resonance simultaneously, the spatial distribution of the total electromagnetic field in the object will in effect be a linear combination of these eigenmodes. Other spatial features, corresponding to eigenmodes further from the resonance, will be relatively suppressed. This obviously resembles the action of periodic crystals, where one prefers to think in terms of plane waves, some sustained and some suppressed by the photonic lattice. By analogy, we may say that the eigenmodes of a homogeneous dielectric resonator form a virtual photonic lattice. When we deal with a finite sample of a periodic crystal the interplay between the two lattices becomes important. This has been realized already in the early work on dynamic X-ray diffraction \cite{15,16}, where the so-called forward-diffracted beam has been observed to emerge right opposite the entry point of the incident beam in the regime of anomalous transmission. Curiously, this work predates not only the modern results on negative refraction by photonic crystal slabs \cite{7,8}, but also the seminal Veselago's paper on negative permittivity and permeability \cite{17}. What we want to show here is that the virtual photonic lattice of a homogeneous dielectric resonator on its own is capable of producing practically the same effects as the real one. All we have to do is to learn to compute, manipulate, and use the natural eigenmodes. 

Although computation of the complete set of eigenmodes and eigenvalues for an object of any reasonably large size is a very challenging numerical problem, we can still make some educated guesses. For example, a lossless dielectric object with the total size much smaller than the medium wavelength does not have any interesting eigenmodes in resonance \cite{13}. In that case, no matter what the eigenmodes actually are, they will have little influence on the spatial distribution of the field. This is also to be expected from a small object, which only slightly scatters the incident light. It does not matter which dimension of the object is small, as it is the total scattering volume that counts. Hence, an object that is relatively large in two dimensions and thin in the third is a weak scatterer and does not have much influence on the incident field. This obvious consideration has a useful application in what follows. 

Consider a simple dielectric parallelepiped and let us vary one of its dimensions, namely, the one transverse to the propagation direction of the incident beam. Figure~1 illustrates the scattering of a TM-polarized Gaussian beam (magnetic field vector is orthogonal to the page; angle of incidence of the beam is 30 degrees) on an object with permittivity $\varepsilon_{0}/\varepsilon=4$. The numerical results were computed using a fully-vectorial three-dimensional code based on the volume integral equation formulation of the electromagnetic scattering problem \cite{12,13,18} and are presented for different values of the transverse thickness. We consider the spatial distribution of the intensity throughout the mid-plane of the object. Some caution has to be exercised in the interpretation of these and similar images. Since we look at a single spatial cross-section, everything we see is by definition effective, i.e., related to the visualized plane only. The light scatters in all directions and its off-plane behaviour is certainly different. Having said that, we observe that, as we expected, the first very thin block does practically nothing to the incident beam letting it through with very little distortion. As the thickness of the block increases we notice the appearance of a standing wave pattern, Figure~1 (top- middle), without much refraction as yet. Judging by the period of the emerging standing wave, the medium wavelength is not at all two times smaller than the free-space wavelength, as one would expect for a medium with $\varepsilon_{0}/\varepsilon=4$. The effective wave speed in a thin block turns out to be faster than in a homogeneous medium with the same parameters. It can be explained by the relatively small volume of the object and the associated lack of scattering. On the other hand, it may be viewed as a manifestation of the so-called wave-guide dispersion, where different spatial modes have different propagation velocities \cite{14}. As the thickness of the block increases, the beam begins to slow down and refract. Simultaneously, however, it splits into two distinct paths, one that refracts more or less normally and the other which bends down, Figure~1 (middle row). This unusual behaviour can easily be explained by resorting to our approximate ideas about the eigenmodes of the dielectric resonator. Indeed, both the horizontal and the vertical standing wave patterns naturally occur due to reflections between the opposite interfaces of the block and most probably belong to the set of eigenmodes. Neither the initial inclined beam nor the expected refracted one, however, has a natural support on the coarse virtual lattice of this thin rectangular resonator. What we observe is the "projection" of a beam onto the dominant spatial modes. In general, the structure of the virtual lattice is expected to be richer than the mentioned vertical and horizontal standing waves. Moreover, as we observe, the lattice period depends on the thickness of the resonator. In particular we notice the appearance of a much finer periodic pattern (see the upper part of image in Figure~1, bottom-middle). Remarkably, it resembles the popular triangular lattice used in photonic crystals and metamaterials, which is now induced by the boundaries of the object rather than imposed by the periodic modulation of permittivity. The effects of this virtual lattice are, however, quite real. As can be seen in Figure~1 (bottom row), when the thickness of the block reaches a little less than half of the medium wavelength, the beam is negatively refracted. In Figure~2 we plot the spatial variation of the intensity along the mid-line of the exit interface for beams of different frequencies incident on a fixed slab. Note that this is not quite the same as our previous example, since the incident field as well as eigenmodes are now changing with frequency. Here too, as the frequency increases the maximum of the transmitted intensity shows progressively stronger positive refraction, followed by the negative refraction at the frequency corresponding to the half of the medium wavelength. This is the case where the effective dispersion makes a prism out of an ordinary transparent resonator which by itself does not have any dispersion on a microscopic level.

 Our simulations demonstrate that the virtual photonic lattice formed by the eigenmodes of a dielectric resonator may produce effects similar to those achieved by real photonic crystals, negative refraction in particular. Unlike its real counterpart, the structure of the virtual lattice can be manipulated by simply varying the transverse dimension, which changes the effective wave speed and the lattice period. Such transversely thin optical devices could, probably, be realized using the standard manufacturing techniques of microelectronics and might provide us with a range of useful optical effects. For example in our simulations we have observed the so-called pendellösung phenomenon, i.e. periodic change between the positive and negative refractions as a function of transverse thickness, which is also known to be a function of the longitudinal dimension \cite{15,16,19}. Our numerical experiments show, however, that the negative refraction by a virtual lattice is a rather narrow-band phenomenon and requires considerable tuning. Also more realistic wider beams require longer distances for the beam splitting to occur (which seems to be essential for the negative refraction in our case) and may therefore be subject to inevitable losses. The main conceptual difference between the real and virtual photonic crystals may be summarized as follows. In an infinite photonic crystal the negative refraction is achieved by effectively blocking the usual direction of propagation of light, so that it has to scatter-off in the ``wrong'' direction at the lower edge of its band-gap. In the virtual finite crystal, we try instead to amplify the ``wrong'' direction by tuning the corresponding modes into resonance, which is generally more difficult as we might simultaneously amplify a whole set of unwanted eigenmodes.  A logical step forward, already considered by some authors for differing reasons [15, 16, 20, 21], is the combination of real and virtual photonic lattices. 
\begin{figure*}[t]
\epsfig{file=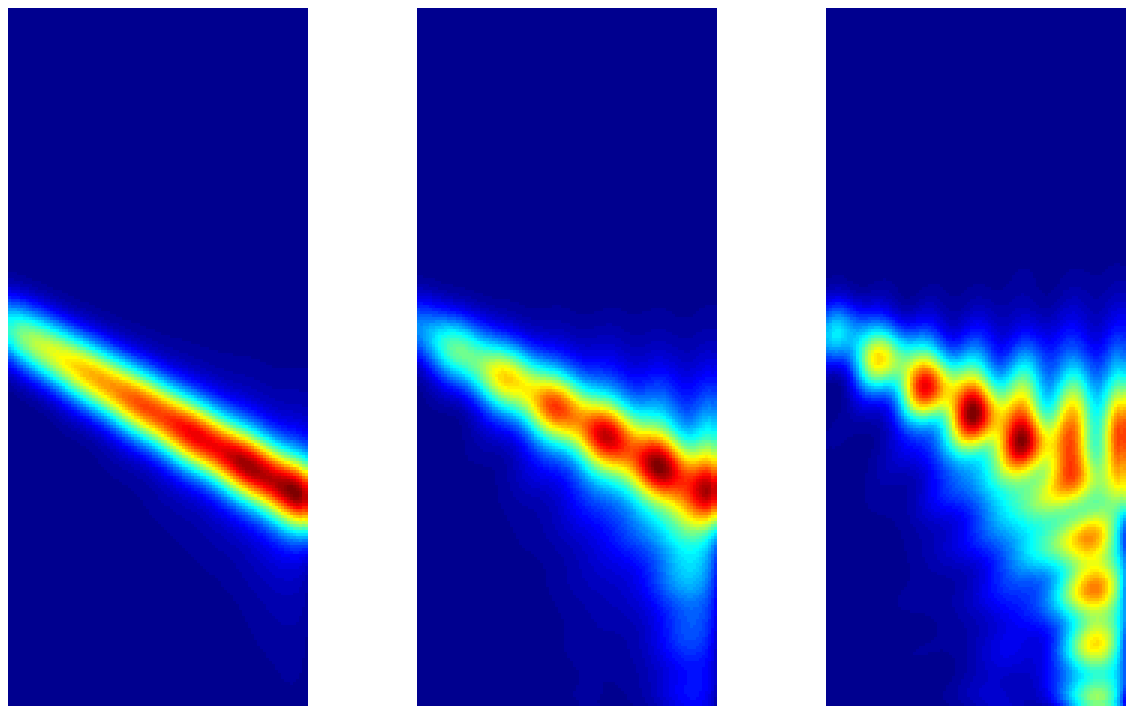,width=0.5\textwidth}
\\ \vspace{3mm}
\epsfig{file=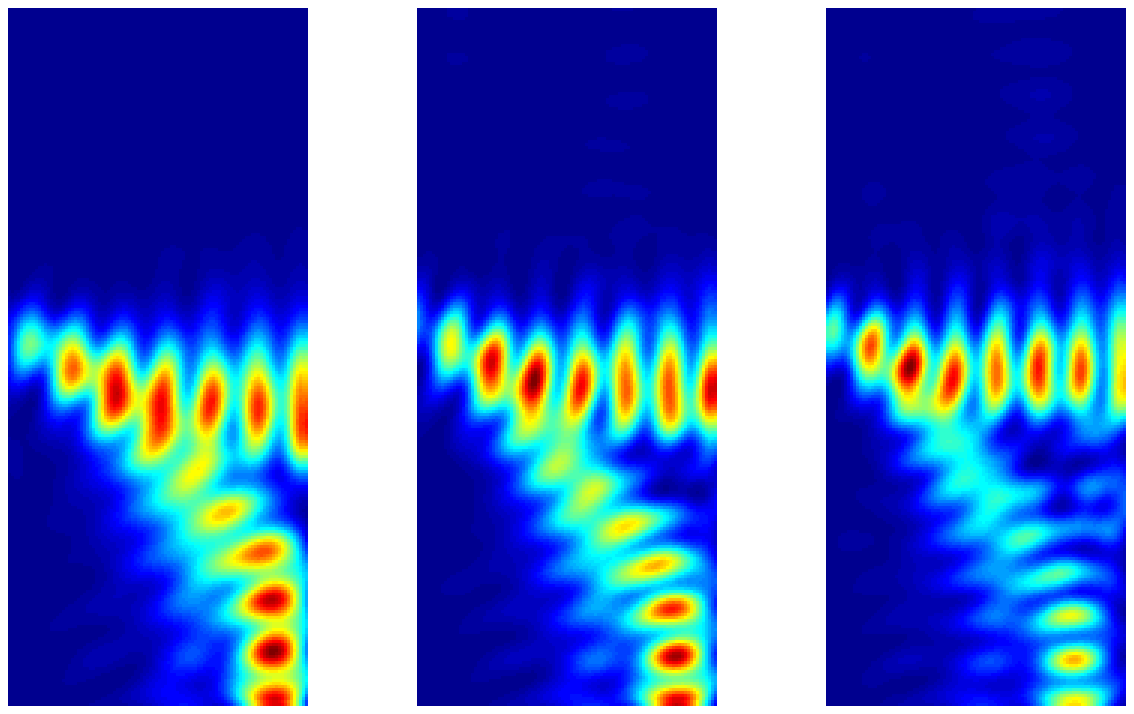,width=0.5\textwidth}
\\ \vspace{3mm}
\epsfig{file=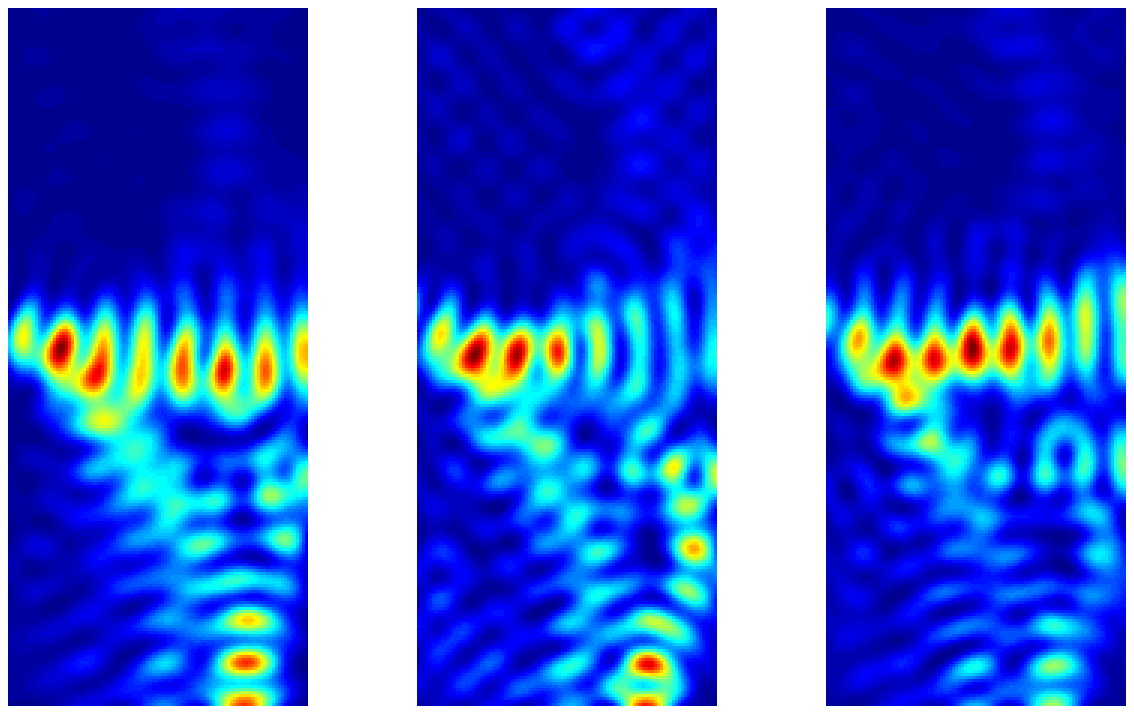,width=0.5\textwidth}
\caption{Refraction of a Gaussian beam by a rectangular three-dimensional dielectric resonator of varying transverse dimension. Resonator parameters: relative permittivity - 4, vertical dimension – 5.825~$\lambda$, horizontal – 2.525~$\lambda$, transverse thickness $d/\lambda$ (left to right, top to bottom): 0.025, 0.05, 0.075, 0.1, 0.125, 0.15, 0.175, 0.2, and 0.225. Gaussian beam is TM-polarized (magnetic field orthogonal to the image plane), incidence angle – 30 deg. Images give field intensity in the cross-sectional mid-plane. One can see the emergent virtual triangular lattice in the upper part of the middle-bottom image.}
\end{figure*}

\begin{figure*}[t]
\epsfig{file=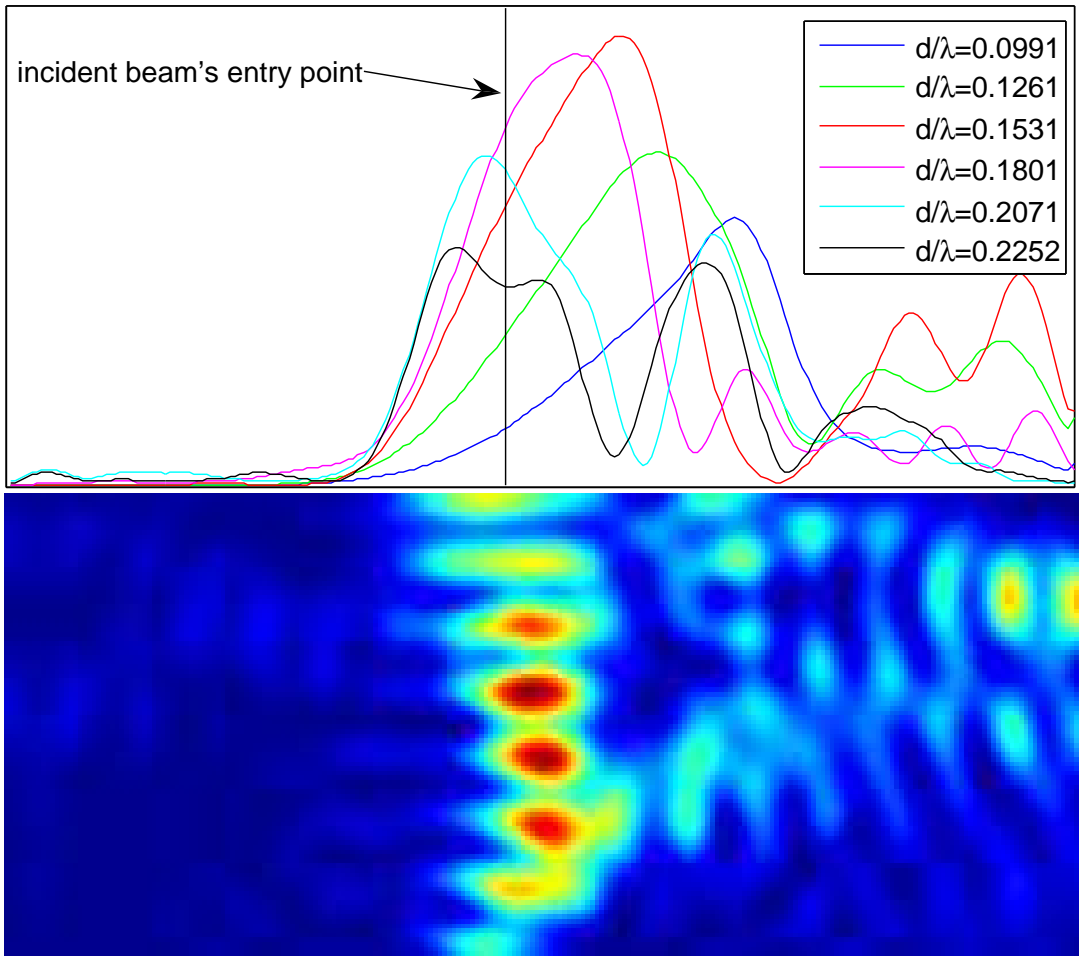,width=0.8\textwidth}
\caption{Dispersion of light of different frequencies by a homogeneous rectangular dielectric resonator – prism effect. Upper image: transmitted intensity along the exit interface for a fixed-size resonator with transverse thickness $d$, illuminated by the Gaussian beam at 30 degrees for different wavelengths. Leftwards shifts of the main maximum indicate dispersion resulting in progressively stronger refraction followed by the negative refraction. Secondary peaks correspond to the incident beam (the beam cross-section is wider than the transverse dimension of the resonator) and standing waves inside the resonator. Lower image: Intensity inside the resonator for $d/\lambda$ = 0.2071.}
\end{figure*}

\end{document}